# Multispectral large-area X-ray imaging enabled by stacked multilayer scintillators

Peng Ran, Lurong Yang, Tingming Jiang, Xuehui Xu, Juan Hui, Yirong Su, Cuifang Kuang, Xu Liu, Yang (Michael) Yang*

State Key Laboratory of Modern Optical Instrumentation,

College of Optical Science and Engineering,

International Research Center for Advanced Photonics,

Zhejiang University, Hangzhou, Zhejiang, China

*Correspondence and requests for materials should be addressed to

Yang (Michael) Yang (yangyang15@zju.edu.cn)

**ABSTRACT**

Conventional energy-integration black-white X-ray imaging lacks spectral information of X-ray photons. Although X-ray spectra (energy) can be distinguished by photon-counting technique typically with CdZnTe detectors, it is very challenging to be applied to large-area flat-panel X-ray imaging (FPXI). Herein, we design multi-layer stacked scintillators of different X-ray absorption capabilities and scintillation spectrums, in this scenario, the X-ray energy can be discriminated by detecting the emission spectra of each scintillator, therefore the multispectral X-ray imaging can be easily obtained by color or multispectral visible-light camera in one single shot of X-ray. To verify this idea, stacked multilayer scintillators based on several emerging metal halides were fabricated in the cost-effective and scalable solution process, and proof-of-concept multi-energy FPXI were experimentally demonstrated. The dual-energy X-ray image of a "bone-muscle" model clearly showed the details that were invisible in conventional energy-integration FPXI. By stacking four layers of specifically designed multilayer scintillators with appropriate thicknesses, a prototype FPXI with four energy channels was realized, proving its extendibility to multispectral or even hyperspectral

X-ray imaging. This study provides a facile and effective strategy to realize energy-resolved flat-panel X-ray imaging.

**INTRODUCTION**

X-ray imaging technology is widely used in medical diagnosis[1-5], security inspection[6-10], and scientific research[11-15]. The image contrast of a typical energy-integration X-ray imager reflects the total X-ray intensity passing through an object.[16-18] The spectral information is absent; thus the image is usually black-white. Materials with similar atomic numbers and densities have little contrast with this imaging mode.[19-25] The attenuation of X-ray is mainly caused by photoelectric absorption and Compton scattering.[26] Both effects are strongly correlated with the X-ray energy, thus the X-ray absorption coefficient of a certain material has a specific dependence on X-ray spectra. Therefore, the energy-resolved X-ray imaging can better identify materials and generate better image contrast by detecting the attenuation of X-ray at different energy channels.[27-29] Currently, the mainstream of energy-resolved X-ray imaging relies on the photon-counting detectors (PCDs).[30-33] In the case of the most commonly used CZT (CdZnTe) crystal PCD[34-37], each X-ray photon produces an electrical pulse with the signal height proportional to the incident X-ray energy. By comparing the pulse heights with certain thresholds, the incoming X-ray photons can be classified by energy.[38] However, the photon-counting technique still has many critical limitations. Firstly, the "pile-up" effect largely reduce or even disable the energy resolvability under high-flux X-rays.[39] Secondly, it is well-known that growing high-quality and large-scale CZT single crystal is very difficult and expensive.[40-44] Thirdly, each PCD unit needs a sophisticated circuit system, thus the pixel pitch is large and the integration density is much lower than the Si-based TFT or CMOS array, making it far from satisfying for large-area FPXI that is currently dominated by scintillator-based imager.[45,46] In summary, although PCDs are starting commercialization in high-end multi-energy Computed Tomography (CT) scanning system, this technique can hardly be directly implemented in the widely-used FPXI, to the best of our knowledge, there is no such multi-energy FPXI yet in the market.

Hence, it is highly desirable to develop energy-resolved FPXI detectors that combine high-performance, cost-effectiveness, and scalability. Recently, many metal halide[47-53] and perovskite[54-59] scintillators with the advantages of spectral tunability and facile preparation have been reported. On this basis, a multilayer stacked scintillator structure was designed and fabricated for large-area multi-energy FPXI. The top layer is designed to be mainly responsive to low-energy X-rays and emitting long-wavelength scintillations; the bottom layer absorbs the high-energy X-rays and produces short-wavelength scintillations. The discrimination of X-ray energy is translated to the detection of emission color of each scintillator, which can be easily realized by color or multispectral camera. Unlike the PCDs that need to scan the object to obtain the image, our method utilizes the matured Si-based image sensor to acquire multi-energy X-ray image in a single shot of X-ray. In this study, the prototype dual and quaternary energy X-ray imaging systems were demonstrated with stacked two and four scintillators, respectively. They demonstrate clear advantages over the conventional FPXI in identifying materials of similar densities.

**RESULTS AND DISCUSSION**

Fig.1a shows the conventional FPXI system. The scintillation intensity represents the integrated X-ray energy passing through the object, which is detected by CMOS or TFT image sensor. Fig.1b shows the principle of the energy-resolved FPXI based on stacked multilayer scintillators. Here we used four stacked layers with four X-ray energy bins as a schematic, and it can certainly be applied to more layers with more energy bins. The scintillator films made of different materials are stacked perpendicular to incident X-ray. From top to bottom, the radioluminescence (RL) wavelength of each scintillator is becoming shorter, which ensures the light emitted by the upper scintillator is transparent to the lower layers. The X-ray absorption coefficient generally decreases with the increase of X-ray energy. In an ideal case, each appropriately selected scintillator with a certain thickness absorbs a certain segment of X-ray and emits specific scintillation. The number of scintillator layers determines the energy channels that can be divided, known as X-ray energy (spectra) resolution. In this schematic (Fig.1b), we illustrate four X-ray energy channels E1~E4. The RL spectra of scintillator R1~R4 should not be overlapped, and they are mainly responsive to the E1~E4, respectively. A multispectral camera is used to collect the luminous image of each scintillator, finally, four separated images are fused into one multi-energy X-ray image. In this scenario, the energy bins can surely be expanded by using a hyperspectral camera with more stacked scintillators of different emission bands. This approach utilizes large-area stacked scintillators and matured Si-based flat-panel sensors to acquire multi-energy FPXI. Compared to the current photon-counting technique, it has a much larger field-of-view (FOV) and the energy-resolved images can be obtained in one shot of X-ray. Besides, in photon-counting mode, if two or more X-ray photons arrive at the detectors simultaneously, the energy information will be misinterpreted, known as the "pile-up" effect. Our proposed method is free of "pile-up", and can work under high-flux X-rays.

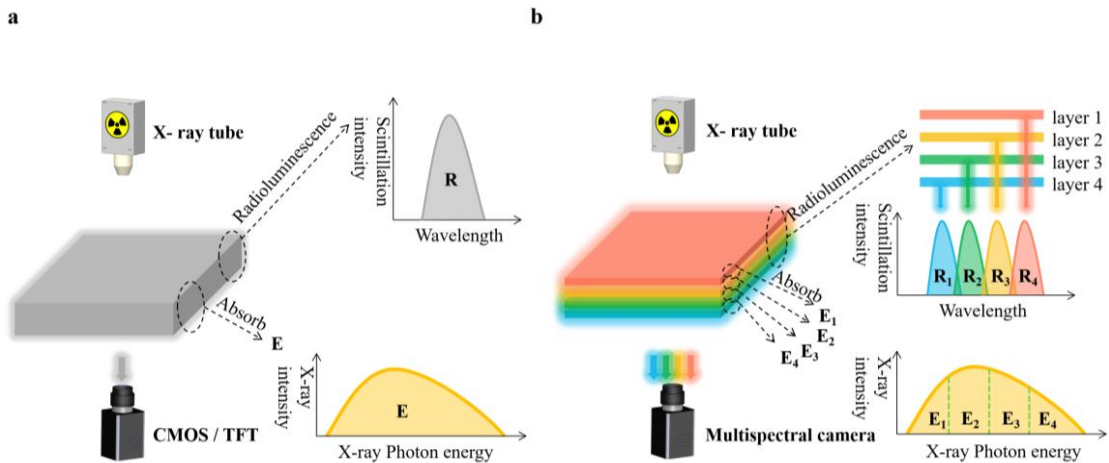

**Fig. 1 | Schematic of energy-integrated and multi-energy X-ray imaging systems.**
**(a)** Schematic of a conventional energy-integrated X-rays imaging system. **(b)** Schematic of a large-area multi-energy flat-panel X-rays imaging system based on stacked multilayer scintillators.

First, the feasibility of dual-energy X-ray imaging is verified by two stacked scintillators based on emerging metal halides. Blue-emission $Cs_3Cu_2I_5$ with high X-ray absorption cross-section is designed as the bottom scintillator, and it exhibits a high yield of 79,000 photons MeV$^{-1}$.[60] $C_4H_{12}NMnCl_3$ is the upper scintillator, and it has a steady-state internal light yield of 50,500 photons MeV$^{-1}$.[61] Fig.2a shows the radioluminescence (RL) spectra of $Cs_3Cu_2I_5$ and $C_4H_{12}NMnCl_3$ powder. The emission peaks of $Cs_3Cu_2I_5$ and $C_4H_{12}NMnCl_3$ are 452 nm, 646 nm respectively, with no overlap in emissions. As shown in Fig.S5, there is also no overlap between the optical absorption of $Cs_3Cu_2I_5$ and the radiation luminescence (RL) of $C_4H_{12}NMnCl_3$, avoiding energy transfer or reabsorption between them. The narrower optical bandgap of the upper scintillator warrants the scintillation from the top layer is transparent to the bottom scintillator. $Cs_3Cu_2I_5$ was synthesized by anti-solvent method, $C_4H_{12}NMnCl_3$ was prepared by the slow-evaporation method, and their X-ray diffraction patterns (Fig.S6) agree with previous litterature.[61,62] Fig.2b shows the X-ray absorption coefficients of $Cs_3Cu_2I_5$, $C_4H_{12}NMnCl_3$, and some other emerging as well as conventional scintillators within the energy range from 10$^{-3}$ to 10 MeV. The absorption coefficient of $Cs_3Cu_2I_5$ is close to the commercial CsI and much larger than that of $C_4H_{12}NMnCl_3$, as shown in the inset of Fig.2b. Fig.2c presents the RL of $Cs_3Cu_2I_5$ and $C_4H_{12}NMnCl_3$, both show good linear dependence on X-ray dose. The scintillator screens were fabricated by mixing them with polymethyl methacrylate (PMMA). The scanning electron microscopy (SEM) images reveal that the scintillator particles are uniformly embedded in the matrix, with relatively good transparency as shown in Fig.S3. As shown in Fig.S4, the scintillation film can be bent and preserved in the air for a long time, due to the protection of PMMA.

According to the above method, $Cs_3Cu_2I_5$ and $C_4H_{12}NMnCl_3$ were made into a series of scintillator films of different thickness as displayed in Fig.S7. Since the X-ray absorption of PMMA is negligible, the equivalent thicknesses of the film can be calculated with following equation:

$$T = \frac{M}{\rho \times S}$$

Where T is the equivalent thickness of the film, M is the mass of the scintillator, $\rho$ is the density, S is the film area. As exhibited in Fig.2d, the light output of the two scintillator films under X-ray increases almost linearly with the thickness, illustrating that the self-absorption is negligible. The inset of Fig.2d showed the photographs of $Cs_3Cu_2I_5$ and $C_4H_{12}NMnCl_3$ films under X-rays, exhibiting intense blue and red emissions. As shown in Fig.2e, with the single $Cs_3Cu_2I_5$ scintillator, the inner structures of a circuit board and a dried fish could be clearly observed under 50 kV X-ray. Similarly, we could acquire X-ray images of a capsule with a spring inside and a chip with multiple pins as shown in Fig.2f by using single $C_4H_{12}NMnCl_3$ scintillator. As shown in Fig.2g, the observable X-ray imaging resolution of $Cs_3Cu_2I_5$ film (11 μm) and $C_4H_{12}NMnCl_3$ film (84 μm) are ~12.0 and ~9.0 lp mm$^{-1}$, respectively. To further quantify the resolution, we utilized the slanted-edge method to obtain the modulation transfer function (MTF) curve by imaging a 1 mm thick aluminum slice with a sharp edge. As shown in Fig.2h, the spatial resolution (@MTF = 0.2) of those scintillators are determined to be 12.2 and 9.1 lp mm$^{-1}$, which are consistent with the observation limit. Those experiments prove that each individual scintillator can produce good-quality X-ray image, laying the foundations for energy-resolved X-ray imaging.

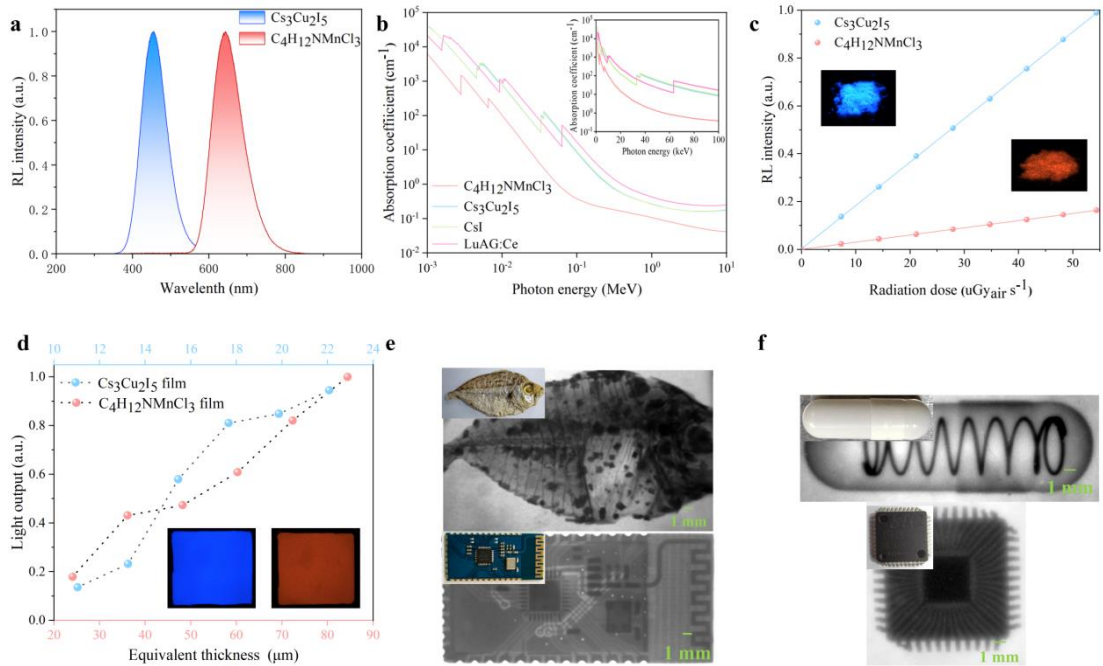

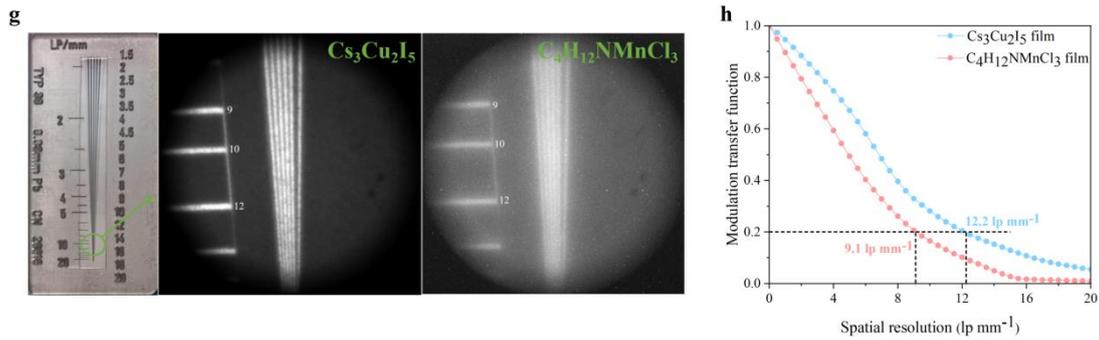

**Fig.2 | Radioluminescence (RL) characterizations and X-ray imaging experiments based on $Cs_3Cu_2I_5$ and $C_4H_{12}NMnCl_3$.**
**(a)** RL spectra of $Cs_3Cu_2I_5$ and $C_4H_{12}NMnCl_3$. **(b)** Absorption coefficients of $Cs_3Cu_2I_5$, $C_4H_{12}NMnCl_3$ and typical CsI and LuAG:Ce as a function of X-ray energy. **(c)** The RL intensity of $Cs_3Cu_2I_5$ and $C_4H_{12}NMnCl_3$ powder under X-ray irradiation with different doses. **(d)** The light output of $Cs_3Cu_2I_5$ and $C_4H_{12}NMnCl_3$ scintillator films of different equivalent thicknesses. Inset is the photograph of two scintillator films under X-ray radiation. **(e)** Photograph of the test objects and their X-ray images by $Cs_3Cu_2I_5$ scintillator film (11 μm). **(f)** Photograph of the test objects and their X-ray image by $C_4H_{12}NMnCl_3$ scintillator film (84 μm). **(g)** Photograph of the standard X-ray test-pattern plate and X-ray images of the test-pattern plate in the partial region. **(h)** Modulation transfer function (MTF) curves of a metallic sharp edge based on individual $C_4H_{12}NMnCl_3$ and $Cs_3Cu_2I_5$ scintillator.

Subsequently, two scintillators of different thicknesses are stacked for dual-energy X-ray imaging. The X-ray absorption efficiencies of different thickness combinations are shown in Fig.3a. It can be straightforwardly seen that the low energy segments of X-rays are mainly absorbed by the upper scintillator, whereas the high-energy parts are absorbed by the below one. This selective absorption characteristic is especially obvious for the combination of 84 μm $C_4H_{12}NMnCl_3$ and 11 μm $Cs_3Cu_2I_5$. Using this combination, we tested its light output in the stacked configuration. As shown in Fig.3b, the light output of the stacked scintillator only slightly decreased in the $C_4H_{12}NMnCl_3$ (upper scintillator) emission band indicating most of the scintillation from upper layer can transmit through the bottom scintillator. The decrease of the stacked scintillators in the $Cs_3Cu_2I_5$ (bottom scintillator) emission band is clearly originating from the X-ray absorption of upper scintillator. The X-ray imaging resolutions produced by blue and red scintillations in the stacked configuration were measured to be 10.8 and 6.8 lp mm$^{-1}$(Fig.3c), which were slightly lower than the cases when they were tested separately. This is due to the slightly increased light scattering. A "bone and muscle" model (Fig.3d) was made from aluminum and cardboard, simulating bone and muscle tissue respectively, because of their similar X-ray absorption coefficients. The holes in aluminum and cardboard (Fig.3e) mimic the defects in bones and muscles, representing the targets for X-ray examination. Dual-energy X-ray imaging of the "bone and muscle" was performed, and compared with conventional energy-integration X-ray imaging based on $Cs_3Cu_2I_5$ and $C_4H_{12}NMnCl_3$ single-layer scintillators. To qualitatively compare their discriminability of bone and muscle defects, we defined two terms to describe the image contrast at a defect site of bone ($K_B$) and muscle ($K_M$). The $K_B$ and $K_M$ are defined as follows:

$$K_B = \frac{I_{B_1} - I_{B_2}}{I_{B_2}} \qquad K_M = \frac{I_{M_1} - I_{M_2}}{I_{M_2}}$$

Where $I_{M1}$ and $I_{M2}$ are the average gray values at muscle defect area $M_1$ and its surrounding area $M_2$ respectively, $I_{B1}$ and $I_{B2}$ are the average gray values at bone defect area $B_1$ and its surrounding area $B_2$ respectively. We varied the thicknesses of $C_4H_{12}NMnCl_3$ and $Cs_3Cu_2I_5$, acquired respective images and calculated the corresponding $K_M$ and $K_B$, under different X-ray tube currents (Fig.3f and Fig.3g). We find that single $Cs_3Cu_2I_5$ scintillator produces high image contrast for bone defect ($K_B$), but low contrast for muscle defect ($K_M$); whereas for single $C_4H_{12}NMnCl_3$, the contrary is the case. This is because the image contrast of low-density muscle is mainly contributed by low-energy X-ray, and the contrast of high-density bone is mainly caused by high-energy X-ray. With stacked scintillators, we will be able to observe both defects in different X-ray energy channels. We screened the thicknesses of bottom and upper scintillators, and found 84 μm $C_4H_{12}NMnCl_3$ plus 11 μm $Cs_3Cu_2I_5$ delivered the best $K_M$ and $K_B$ (Fig.3f and Fig.3g), which agrees with our calculations in Fig.3a. Fig.3h showed the optimized $K_B$ and $K_M$ with stacked $Cs_3Cu_2I_5$ (11 μm) and $C_4H_{12}NMnCl_3$ (84 μm), at different X-ray illumination intensities. Finally, as shown in Fig.3i, only the stacked scintillator can give clear image contrast in both defects of bone and muscle, the mere $C_4H_{12}NMnCl_3$ or $Cs_3Cu_2I_5$ scintillator can only display either muscle or bone defect. This experiment demonstrates clear advantage of our proposed dual-energy FPXI over conventional energy-integration FPXI.

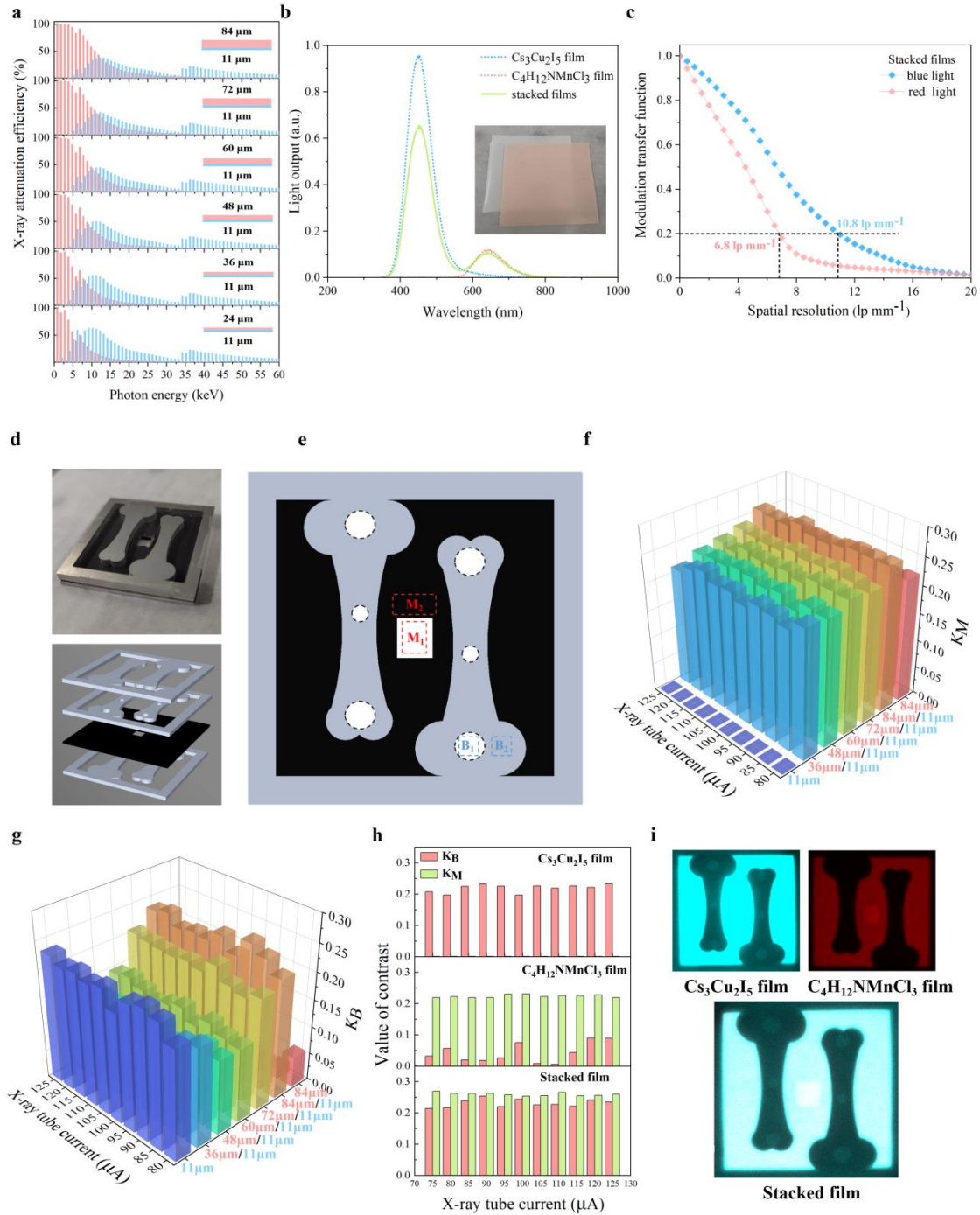

**Fig. 3 | Dual energy X-ray imaging enabled by dual-layer stacked scintillators.**
**(a)** The X-ray absorption efficiencies of different film thickness combinations of top $C_4H_{12}NMnCl_3$ and bottom $Cs_3Cu_2I_5$. **(b)** The light output of stacked scintillators and separated scintillators under X-ray. **(c)** Modulation transfer function (MTF) curves of stacked scintillators, tested by the slanted-edge method. **(d)** Structure drawing of a bone - muscle test model. **(e)** Schematic of defects in bone and muscle areas. **(f)** The X-ray image contrast of muscle defects ($K_M$) of different scintillator thickness combinations. **(g)** The X-ray image contrast of bone defects ($K_B$) with different scintillator thickness combinations **(h)** The $K_B$ and $K_M$ values of X-ray images of separated $Cs_3Cu_2I_5$ $C_4H_{12}NMnCl_3$ and stacked scintillators (84 μm $C_4H_{12}NMnCl_3$ /11 μm $Cs_3Cu_2I_5$). **(i)** Dual-energy X-ray images of the "bone and muscle" model by stacked scintillators (84 μm $C_4H_{12}NMnCl_3$ /11 μm $Cs_3Cu_2I_5$).

The energy channels can certainly be extended by stacking more scintillators, and the strong tunability of metal halides makes it feasible. Here, a four-layer tandem scintillator made of distinct metal halides was constructed to demonstrate energy-resolved X-ray imaging. The four scintillators from top to bottom are $FAPbI_3$, $C_4H_{12}NMnCl_3$, $(C_8H_{20}N)_2MnBr_4$, and $Cs_3Cu_2I_5$, respectively. From top to bottom, the scintillation spectra is not overlapped and blueshifts, ensuring that the emission from the upper layer is fully transparent to the below layers (Fig.4a). Fig.4b presents the X-ray absorption coefficient of the four materials at the energy range from 1 to 100 keV. The maximum of our X-ray source was 60 keV, thus we divided four energy channels centered at 5,15,30 and 50 keV. Fig.S8 shows the attenuation efficiency of the four scintillators with different thicknesses. According to X-ray attenuation law, each scintillator of appropriate thickness can be designed to absorb the majority of X-ray of one energy segment. As exhibited in Fig.4c, the scintillator thickness of the four layers was designed to be 3, 300, 400, and 50 μm, respectively, they are mainly responsible for the X-ray segments at 5,15,30 and 50 keV. As shown in Fig.4e, 55.2% of the light emitted by the first layer comes from the absorption of 5 keV X-ray, 53.1% scintillation of the second layer comes from 15 keV X-ray, 50.2% scintillation of the third layer comes from 30 keV X-ray and 78.4% scintillation of the fourth layer comes from 50 keV X-ray. On basis of those results, the multi-layer scintillator is used to conduct energy-resolved X-ray imaging of the same "bone and muscle" model, and four X-ray images of different energy channels are obtained as shown in Fig.4d. It proves the feasibility to realize multispectral of even hyperspectral large-area X-ray imaging based on stacked scintillators, especially given the fast developments of hyperspectral camera and metal halide scintillators.

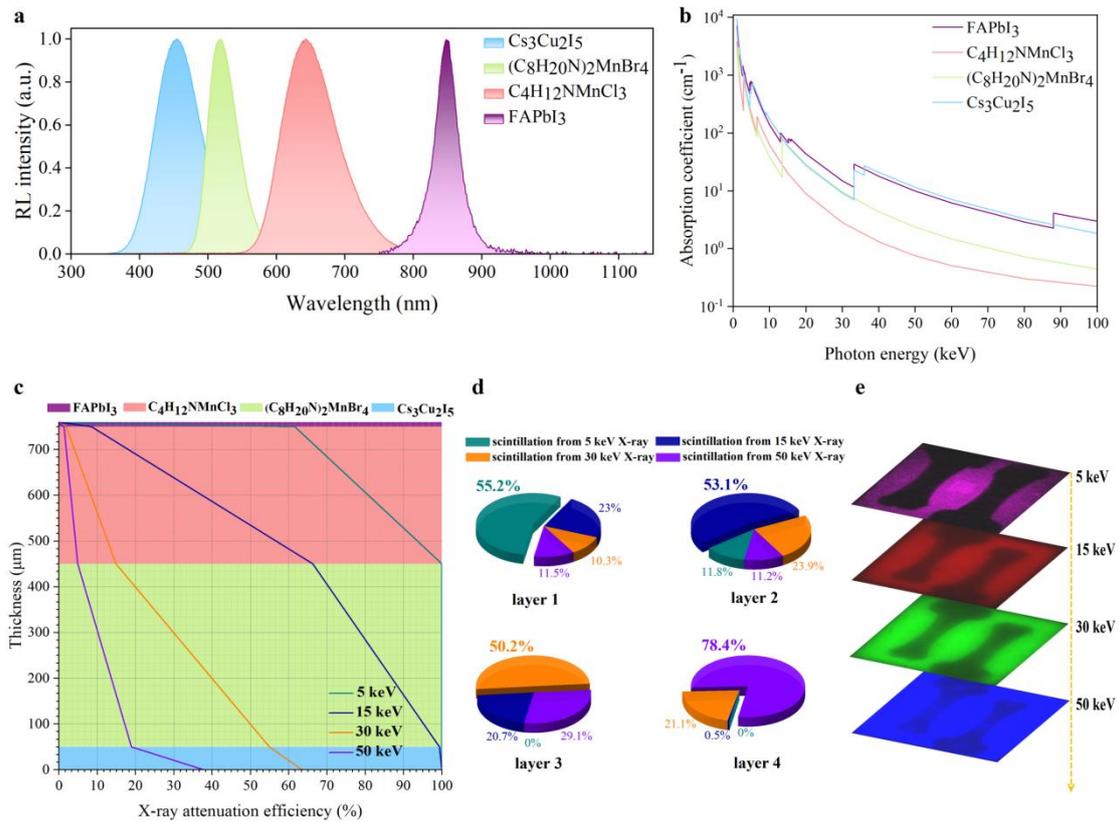

**Fig. 4 | Multi-energy X-ray imaging based on four-layer stacked scintillator films.**
The radiation luminescence (RL) spectra of $FAPbI_3$, $C_4H_{12}NMnCl_3$, $(C_8H_{20}N)_2MnBr_4$ and $Cs_3Cu_2I_5$ scintillators, from the top to the bottom. **(b)** X-ray absorption coefficients of $FAPbI_3$, $C_4H_{12}NMnCl_3$, $(C_8H_{20}N)_2MnBr_4$ and $Cs_3Cu_2I_5$ from 1 keV to 100 keV. **(c)** The thickness-absorption efficiency relation of the four scintillators at different energy X-ray. **(d)** The proportion of the four energy X-rays contributing to the light emitted by each scintillator. **(e)** Multi-energy X-ray images at four energy channels at 5, 15, 30, and 50 keV.

## CONCLUSION

In conclusion, we realized a new strategy for multispectral flat-panel X-ray imaging by stacked multilayer scintillators based on a low-cost and scalable approach. Compared with existing mainstream energy-resolved detectors (such as PCDs), our scintillator-based method is fully compatible with the matured visible-light image sensors, enabling convenient realization of multispectral FPXI in a single shot of X-ray. Since it doesn't involve photon counting, , it is free of "pile-up" and can be used under high-flux X-rays. The proof-of-concept prototype dual and four-energy X-ray imaging were demonstrated with clear advantage over conventional energy-integration flat-panel X-ray imaging.    Considering the rapid developments of hyperspectral camera and deep-learning enabled spectrum-sensing algorithm, our results unfold the huge potentials of multi or hyperspectral X-ray imaging technique based on stacked multilayer scintillators.

ACKNOWLEDGEMENTS

Y.(M.)Y. acknowledges funds received from the National Key Research and Development Program of China (2017YFA0207700), and the Natural Science Foundation of China (62074136, 61804134, 61874096).

**Author Contributions Statement**

Y.Y.(M) conceived the idea and supervised the experiments. P.R. performed all the X-ray imaging related experiments, analyzed the data and wrote the first draft of the manuscript. P.R., L.Y, T.J., X.X. J.H. and Y.S. prepared the scintillator materials used in this study. X.L. and C.K. provided insightful discussions. Y.Y. (M) majorly revised the manuscript with inputs from all the authors.

**Competing Interests Statement:**

The authors declare no competing interests.